\documentstyle[preprint,revtex]{aps}     
\renewcommand{\baselinestretch}{1.346}   
\begin{document}

\draft

\begin{title}
Dynamical decoupling and Kac-Moody current\\
representation in multicomponent integrable systems
\end{title}

\author{J.M.P. Carmelo and A.H. Castro Neto}
\begin{instit}
Department of Physics, University of Illinois at Urbana-Champaign\\
1110 West Green Street, Urbana, Illinois 61801-3080
\end{instit}
\receipt{September 1992}

\vspace{0.8cm}                           
\begin{abstract}
The conformal invariant character of $\nu$-multicomponent integrable
systems (with $\nu$ branches of gapless excitations) is described
from the point of view of the response to curvature of the
two-dimensional space. The $\nu\times\nu$ elements of the dressed
charge matrix are shown to be transition matrix elements of the zero
($\mu =0$) components of the diagonal generators of $\nu$ independent
Kac-Moody algebras (Cartan currents). The dynamical decoupling which
occurs in these systems is characterized in terms of the conductivities
associated with the $\mu = 1$ components of the Cartan currents.
\end{abstract}
\vspace{0.7cm}                           
\renewcommand{\baselinestretch}{1.656}   

\pacs{PACS numbers: 05.70. Jk, 64.60. Fr, 72.15. Nj, 67.90. +z}

\narrowtext
Conformal invariance has been shown to be particularly powerful
in two dimensions \cite{Blote,Houches}. Critical theories are
parametrized by the conformal anomaly $c$, which can both be
extracted from (i) the finite-size-scaling behavior of the
ground-state energy; and from (ii) the stress tensor-tensor
correlation function associated with the response to curvature
of the two-dimensional space \cite{Blote}. On the one hand,
however, the method (ii) has not been applied directly to
$\nu$- multicomponent ($\nu>1$) models solvable by Bethe ansatz
\cite{Izergin,Frahm}, since conformal field theory refers to
Lorentz-invariant systems only. The 1-D Hubbard and supersymmetric
$t-J$ models \cite{Frahm}, and some of the multicomponent systems
with $(1/r^2)$ long-range interaction \cite{Kawa} are examples
of such systems. On the other hand, and although the gapless
excitations of these systems have different ``light'' velocities,
results obtained in finite-size-scaling studies have suggested that
each gapless excitation corresponds to one Virasoro algebra of
conformal anomaly $c=1$ and that the complete critical theory is
given as a direct product of $\nu$ Virasoro algebras \cite{Frahm}.

In this Letter we study the conformal-invariant character of
the above multicomponent systems from the point of view (ii).
Their Bethe-ansatz solutions have a universal character
\cite{Izergin,Frahm,Carmelo92b} which allows the present
general study (the details of our calculations will be presented
elsewhere \cite{Carmelo92b}). Although we consider the ``phases''
of lowest symmetry $[U(1)]^{\nu}$, our results provide full
information about the higher symmetry phases. The energy-momentum
tensor decouples in $\nu$ new tensors which act on orthogonal
sub-Hilbert spaces. Each gapless excitation branch corresponds
to independent Minkowski spaces (with common space and time
but different ``light'' velocities). The Lorentz-invariance in each
of these $\nu$ spaces is associated with $\nu$ independent
Virasoro algebras and the $\nu$ related Kac-Moody algebras
\cite{Houches,Carmelo92b}. We denote the diagonal generators
of the later algebras, which define Cartan sub-algebras, by Cartan
currents. Furthermore, we find a close connection between the dressed
charge matrix $Z^1$ of Korepin \cite{Izergin,Frahm,Carmelo92b}
and the generators of the Kac-Moody algebras -- the elements of $Z^1$
(which determine the non-classical critical exponents \cite{Frahm})
are shown to be the $\nu\times\nu$ different non-vanishing
matrix elements of the $\nu$ Cartan-current zero-components
in the basis of the Hamiltonian eigenstates of lowest energy. Another
central result is the characterization in terms of the properties of
the conductivity stiffnesses associated with the Cartan currents of
the dynamical decoupling which is shown to occur in these systems.

We consider the sector of parameter space which corresponds to
values of the fields (magnetic fields, chemical potentials, and
other generalized fields associated with the conserved quantities
below $(1)$ and $(2)$) where the symmetry is $[U(1)]^{\nu}$.
For example, in the case of the Hubbard-chain Landau-Luttinger
phase (HLLP) the symmetry is $[U(1)]^2$ (densities $n\neq 1$ and
magnetic fields $H>0$), becomes $U(1)\times SU(2)$ at $H=0$
\cite{Carmelo92b,Carmelo92}, and at zero chemical potential and
$H=0$ is $SO(4)$ \cite{Essler}. The advantages of considering the
$[U(1)]^{\nu}$ phases are: (I) all the gapless branches of
low-lying eigenstates are generated by pseudoparticle-pseudohole
processes, i.e. these systems are 1-D Landau liquids
\cite{Carmelo92}; (II) taking the limiting values of the fields
which characterize the higher symmetry phases in the expressions
for the $[U(1)]^{\nu}$ phases provides the correct values for the
conformal dimensions and physical quantities in the former phases
\cite{Carmelo92b}. In the $[U(1)]^{\nu}$ phases the low-energy
physics is described by $\nu$ liquids of interacting $\alpha$-
pseudoparticles \cite{Carmelo92}. These pseudoparticles are the
transport carriers, coupling to external forces
\cite{Carmelo92b,Carmelo92}. In addition to the $\nu$ pseudoparticle
quantum numbers $\alpha$ (one for each branch of pseudoparticles:
for example, in the HLLP we have $\alpha$=c,s, which refers to the
c- and s-pseudoparticles \cite{Carmelo92} -- it is not entirely
correct to identify c and s with charge and spin, respectively, as
shown below), the pseudoparticles are characterized by an arbitrary
phase and the pseudomomentum quantum number $q$. In the
thermodynamic limit the low-energy excitations are described by
$\nu$ pseudomomentum distributions of fermionic character
$N_{\alpha}(q)=\Theta (q_{F\alpha}-|q|)+\delta_{\alpha} (q)$,
where $|q|<q_{\alpha}$. The limits of the ``Brillouin zones''
$\pm q_{\alpha}$ are determined by the restrictions of the
Bethe-ansatz numbers \cite{Carmelo92b,Carmelo92}. Here
$N_{\alpha}^0(q)=\Theta (q_{F\alpha}-|q|)$ refers to the ground state
and $\delta_{\alpha} (q)$ is the $\alpha$- pseudomomentum
deviation which describes $\alpha $- pseudoparticle-pseudohole
processes (or adding and removal of $\alpha$- pseudoparticles
\cite{Carmelo92b,Carmelo92}). The $\nu $ values of pseudo-Fermi momenta
$q_{F\alpha}$ are determined by the $\nu $ conserved quantities:

\begin{equation}
N_{\gamma (\alpha)} \equiv {N_a\over {2\pi}}
\int_{-q_{\alpha}}^{q_{\alpha}}dq N_{\alpha }(q) \, ,
\end{equation}
being given by $q_{F\alpha}=\pi N_{\gamma (\alpha)}/N_a\leq q_{\alpha}$
($N_a\rightarrow \infty$ -- number of lattice sites). Note that $\gamma
(\alpha)$ are {\it particle quantum numbers} (charge, spin, etc.).
A crucial point is that while in the $[U(1)]^{\nu}$ phases
$q_{F\alpha}<q_{\alpha}$ for all $\nu$ branches, in higher symmetry
phases $q_{F\alpha}=q_{\alpha}$ for one or several branches,
and therefore the corresponding gapless excitations are not of
pseudoparticle-pseudohole type \cite{Carmelo92b}. When, upon
varying the corresponding field, one of the momenta
$q_{F\alpha}\rightarrow q_{\alpha}$, either occurs a phase
transition (opening of a gap), or the collapsing branch is replaced
by a new gapless non-pseudoparticle-pseudohole branch which
corresponds to a higher symmetry \cite{Carmelo92b}
($q_{F\alpha}=q_{\alpha}$ for all $\nu$ branches in the
highest symmetry case). In the HLLP, $\gamma (c)=\rho $
($\rho$ -- charge) and $\gamma (s)=\downarrow $ (down spin), the
conserved quantities $(1)$ being the total number of electrons
$N_{\rho}=N$ and of down-spin electrons $N_{\downarrow}$, respectively
(and then $q_{Fc}=2k_F$ and $q_{Fs}=k_{F\downarrow}$. In addition,
$q_{c}=\pi$ and $q_{s}=k_{F\uparrow}$ \cite{Carmelo92}). In
non-interacting systems always $\gamma (\alpha )=\alpha $ (for
the ($U=0$) Hubbard-chain non-interacting phase (HNIP), where the
excitation spectrum is generated by real electron-hole processes
\cite{Carmelo92}, $\gamma (\sigma )=\sigma=\uparrow ,\downarrow$).
Let us also consider general conserved quantities of the form

\begin{equation}
N_{\vartheta} = \sum_{\alpha '} g^{\vartheta }_{\alpha '}
N_{\gamma (\alpha ')} \, , \hspace{1.5cm}
(g^{\gamma (\alpha )}_{\alpha '} = \delta_{\alpha , \alpha '}
\hspace{0.5cm} if \hspace{0.5cm} \vartheta = \gamma (\alpha )) \, ,
\end{equation}
where $g^{\vartheta }_{\alpha '}$ are integers \cite{Carmelo92b}
and $\vartheta $ is a particle quantum number associated with the
conserved quantity (in the HLLP we may choose $\vartheta=
\sigma_z \, ,\uparrow $, where for spin $g_c^{\sigma_z }=1$,
$g_s^{\sigma_z }=-2$, and for up spin $g_c^{\uparrow }=1$,
$g_s^{\uparrow }=-1$).

The energy spectrum of 1-D Landau liquids is a functional of the
pseudomomentum deviations \cite{Carmelo92b,Carmelo92}. The first
functional derivatives define the $\nu$ pseudoparticle bands
$\epsilon_{\alpha} (q)$ and the group velocities $v_{\alpha}(q)
\equiv d\epsilon_{\alpha}(q)/dq$. In particular, the $\nu$ velocities
$v_{\alpha}\equiv v_{\alpha}(q_{F\alpha})$ play a determinant role
at the critical point, being the ``light'' velocities which appear
in the conformal-invariant expressions
\cite{Frahm,Carmelo92b,Carmelo92}. We distinguish the pseudoparticle
single-pair and multipair eigenstates. The excitation energy and
momentum of the former states are given by $\omega_{\alpha}(q,k)=
\epsilon_{\alpha}(q+k)-\epsilon_{\alpha}(q)$ and $k$, respectively
(here $|q|<q_{F\alpha }$ and $q_{F\alpha }<|q+k|<q_{\alpha }$). We
denote the $\alpha$ single-pair eigenstates by $|q,k;\alpha\rangle$.
For later use we introduce the reduced Hilbert space ${\cal H}_0$
spanned by the ground state $|0\rangle $ and the gapless
pseudoparticle single-pair eigenstates of lowest energy $\omega
\rightarrow 0$ (which constitute a complete orthogonal basis
in ${\cal H}_0$). These $4\nu$ states have exclusively excitation
momenta $k\rightarrow 0^{\pm}$ ($2\nu$ states $\lim_{k\to 0^{\pm} }
|\pm q_{F\alpha}^{\mp },k;\alpha\rangle $ of energy $\omega=\pm
v_{\alpha}k\rightarrow 0$) and $k\rightarrow \pm 2q_{F\alpha}$
($2\nu$ states $\lim_{k\to \pm 2q_{F\alpha} }|\mp
q_{F\alpha}^{\pm },k;\alpha\rangle $ of energy $\omega=\pm
v_{\alpha}[k\mp 2q_{F\alpha }]\rightarrow 0$).The excitation energy
of the multipair eigenstates involving a finite density of
pseudoparticles includes the $f$ functions and other higher order
functional amplitudes \cite{Carmelo92b,Carmelo92}. These are
combinations of the velocities and phase shifts
$\Phi_{\alpha\alpha '}(q,q ')$, which measure the phase shift of
the pseudoparticle $\alpha '$ due to a binary forward-scattering
collision with the pseudoparticle $\alpha $, and define the
$S$-matrix for elementary excitations \cite{Carmelo92b,Carmelo92}.
At the critical point the phase shifts $\Phi_{\alpha\alpha
'}^{\pm }\equiv \Phi_{\alpha\alpha '}(q_{F\alpha},\pm q_{F\alpha
'})$ play the relevant role. There we only need to consider the
sub-Hilbert spaces including the following types of low-energy
eigenstates \cite{Carmelo92b}, (A) $\alpha$ eigenstates associated
with an infinitesimal change $\delta N_{\gamma (\alpha)}$ in the
$\nu$ conserved quantities $(1)$ (these eigenstates imply an
infinitesimal change in $q_{F\alpha}$, i.e. adding or removal of
a small density of $\alpha $- pseudoparticles); (B) $\alpha$
single-pair and multipair eigenstates involving the transfer of a
small density of $\alpha$- pseudoparticles from the pseudo-Fermi
point $\pm q_{F\alpha}$ to $\mp q_{F\alpha}$, $D_{\alpha}$
denoting the number of $\alpha$- pseudoparticles transfered; and
(C) $\alpha$ single-pair and multipair eigenstates around
$\pm q_{F\alpha}$, the number of particle-hole processes of this
kind being denoted by $N_{\alpha}^{\pm}$ (note that
${\cal H}_0\subset $ (0)$\oplus $(B)$\oplus $(C), where (0) refers
to $|0\rangle $). Introducing in the energy functional the $\nu$
pseudomomentum deviations $\delta_{\alpha }(q)$ representative of
these excitations \cite{Carmelo92b}, we obtain the excitation
spectrum $E-E_0$ for the eigenstates belonging the sub-Hilbert
spaces (A)-(C). The density of $\alpha$- pseudoparticles which
contribute to the excitations (A)-(C) is given by the ratio of the
number of pseudoparticles involved and $N_{\gamma (\alpha)}$, and
is assumed to be finite but small. To second order in the density
of pseudoparticles we find in unities of $2\pi/N_a$ (and
$N_a\rightarrow \infty $), $E-E_0=\sum_{\alpha} v_{\alpha}
[h_{\alpha}^{+}+h_{\alpha}^{-}+N_{\alpha}^{+}+N_{\alpha}^{-}]$,
where the $h_{\alpha}^{\pm }$ terms and the $N_{\alpha}^{\pm}$
terms are generated by the excitations (A),(B) and the excitations
(C), respectively. Here $h_{\alpha}^{\pm } \equiv {1\over 2}
[\sum_{\alpha '}\xi_{\alpha \alpha '}^1 D_{\alpha '} \pm \sum_{\alpha
'}\xi_{\alpha \alpha '}^0 (\delta N_{\gamma (\alpha ')}/2)]^2$,
$\xi_{\alpha \alpha '}^1=\delta_{\alpha ,\alpha '}+ \Phi_{\alpha\alpha
'}^{+}-\Phi_{\alpha\alpha '}^{-}$, and $\xi_{\alpha \alpha '}^0=
\delta_{\alpha ,\alpha '}+\Phi_{\alpha\alpha '}^{+}+\Phi_{\alpha\alpha
'}^{-}$\cite{Carmelo92b}. At the critical point only this second
order (A)-(C) excitation spectrum is relevant (the pseudomomentum
functional describes, however, a lager excitation region
\cite{Carmelo92b}). The use of the finite-size-scaling method leads
precisely to this energy spectrum \cite{Frahm}, with $h_{\alpha}^{\pm
}$ being the conformal dimensions of the primary fields and the
$\nu\times\nu$ coefficients $\xi_{\alpha \alpha '}^1$ ($\xi_{\alpha
\alpha '}^0$) constituting the matrix $Z^1$ ($Z^0=([Z^1]^{-1})^T$)
\cite{Carmelo}. The fact that we obtain the same result in the case
of the infinite system by looking at the relevant low-energy
eigenstates will be shown to follow from the Lorentz invariance
in each sector $\alpha$. This proves that the Frahm and Korepin
formula for the conformal dimensions \cite{Frahm} is correct.
The ``highest weight states'' (HWS) \cite{Houches,Carmelo92b}
are fully determined by the pseudoparticle eigenstates of
(A) and (B). In (C) the $\alpha$ multipair eigenstates can be
described as a direct product of $\alpha$ single-pair eigenstates,
as in a non-interacting system. Therefore the energy is additive in
the numbers $N_{\alpha}^{\pm }$, i.e. the $\alpha$ branches are
fully decoupled in the case of the excitations belonging to the
space (C) \cite{Carmelo92b}. It follows that (C) decouples into
a set of smaller sub-Hilbert spaces corresponding to each of the
$\alpha$ branches. Due to the orthogonality of these $\alpha$ spaces,
we can uniquely define the projections of the Hamiltonian and
momentum operators in each of them. The same holds true for the
energy-momentum tensor $T_{\mu \nu}$, which  when acting on (C) can
be uniquely decoupled as $T_{\mu \nu}=\sum_{\alpha} T_{\mu
\nu}^{(\alpha)}$, where $T_{\mu \nu}^{(\alpha)}$ is the projection
of $T_{\mu \nu}$ in the sub-Hilbert space $\alpha$ of (C)
\cite{Carmelo92b}. An important point is that at constant values of
$N_{\gamma (\alpha )}$ the ground state fluctuations of the finite
system (associated with the response to curving of the space)
correspond, in the infinite system, to low-momentum and low-frequency
excitations of the type (C). Therefore, it is straightforward to
show that in complex coordinates the stress tensor-tensor correlation
function may be written as \cite{Carmelo92b} $\langle T(x_0,x_1)
T(x_0 ',x_1 ')\rangle=\sum_{\alpha ,\alpha '} \langle T(x_0,x_1)^{(
\alpha )}T(x_0 ',x_1 ')^{(\alpha ')} \rangle$, where $\langle
T(x_0,x_1)^{(\alpha)}T(x_0 ',x_1 ')^{(\alpha ')} \rangle =
\delta_{\alpha ,\alpha '}/\{2[((x_0-x_0 ') v_{\alpha})^2+(x_1-x_1
')^2]^2\}$. On the other hand, since the $\alpha$ sub-Hilbert spaces
are orthogonal, each tensor $T_{\mu \nu}^{(\alpha)}corresponds to
an independent
Minkowski space with ``light'' velocity $v_{\alpha}$. Therefore,
we can in each of these independent spaces consider $v_{\alpha}
=1$, what leads to

\begin{equation}
\langle T(x)^{(\alpha)} T(x ')^{(\alpha ')}
\rangle = \delta_{\alpha ,\alpha '}/ [2(x - x ')^4] \, .
\end{equation}
This confirms that the present multicomponent systems correspond
to a direct sum of conformal theories, each possessing a central
charge equal to one, in agreement with the predictions of Frahm
and Korepin \cite{Frahm}. Each tensor $T_{\mu \nu}^{(\alpha)}$
is associated with a Virasoro algebra and the corresponding set of
generators $L_{n}^{(\alpha)}$ and $\bar{L}_{n}^{(\alpha)}$
\cite{Houches,Carmelo92b}. On the one hand, the ``tower'' of states
associated with the terms $N_{\alpha}^{+}+N_{\alpha}^{-}$ of the
energy spectrum can be constructed by the action of the generators
of the Virasoro algebras on the HWS. On the other hand, we have
shown that the ``tower'' terms are associated (in the infinite
system) with the $\alpha$ multipair eigenstates (C). It follows
that the generators $L_{n}^{(\alpha)}$ and $\bar{L}_{n}^{(\alpha)}$
anihilate ($n>0$) and create ($n<0$) $\alpha$-
pseudoparticle-pseudohole pairs (C) \cite{Carmelo92b}. Rather than
refering to particles, the generators of the Virasoro algebra
$\alpha$ refer to the $\alpha $- pseudoparticles. In turn,
the $\nu $ currents $J_{\mu}^{\gamma (\alpha )}$ associated with
the conserved quantities $(1)$ are the diagonal generators of
the $\nu $ Kac-Moody algebras connected to these $\nu$ Virasoro
algebras, and are naturally expressed in terms of particle
operators \cite{Carmelo92b}. The Lorentz invariance associated
with each sub-Hilbert space $\alpha $ (of (C)) implies
the fact that the energy corrections of the finite system are
equivalent to the eigenstate spectrum for the pseudoparticle
excitations considered above. It also implies that the thermal
excitation energy of the infinite system at low temperatures
equals the energy corrections of the finite system, as in the
conformal-invariant single-component systems \cite{Houches}.

In order to prove the close connection of the matrix $Z^1$ to
the Cartan currents we start by considering the $\mu =0$ components
of the general currents $J_{\mu }^{\vartheta }$ associated with
the conserved quantities $(2)$. For example, in the case of
electronic systems $J_0^{\sigma }(k)\equiv
\sum_{\vec{k}'}c^{\dag }_{\vec{k}+\vec{k}',\sigma}c_{\vec{k}',
\sigma}$ for $\vartheta = \sigma = \uparrow , \downarrow$ and
$J_0^{\uparrow }(k) \pm J_0^{\downarrow }(k)$ gives $J_0^{\rho }(k)$
and $J_0^{\sigma_z }(k)$ (for $+$ and $-$, respectively). The matrix
which represents the operator $J_0^{\vartheta }(k)$ in the
$\omega\rightarrow 0$ reduced Hilbert space ${\cal H}_0$  vanishes
for finite values of $k$ other than $k\rightarrow 0^{\pm}$ and
$k\rightarrow \pm 2q_{F\alpha}$. In these limits it is traceless
and symmetric \cite{Carmelo92b} -- the only non-vanishing transition
matrix elements are the ones coupling $|0\rangle $ to the single-pair
eigenstates. For $k\rightarrow 0^{\pm}$ the ground state $|0\rangle $
and the $\nu$ states ($\pm$ -- alternative) $\lim_{k\to 0^{\pm} }
|\pm q_{F\alpha}^{\mp },k;\alpha\rangle $ constitute a complete basis,
while for each of the $\nu$ pairs of momenta $k\rightarrow \pm
2q_{F\alpha}$ it is constituted by $|0\rangle $ and the state
$\lim_{k\to \pm 2q_{F\alpha} } |\mp q_{F\alpha}^{\pm },k;
\alpha\rangle $. In the former case, out of the $(1+\nu )\times
(1+\nu )$ corresponding matrix elements only $2\nu$ are non-vanishing,
and out of these there are $\nu$ different elements $\langle
\alpha | J_0^{\vartheta }(k_{\to 0})|0\rangle \equiv \lim_{k\to
0^{\pm} } \langle \pm q_{F\alpha}^{\mp },k;\alpha | J_0^{\vartheta
}(k) |0\rangle $. In turn, in the case $k\rightarrow \pm
2q_{F\alpha}$, out of the $\nu\times (2\times 2)$ elements of the
corresponding $\nu$ matrices also only $\nu$ are finite and different,
$\langle \alpha | J_0^{\vartheta }(k_{\to 2q_{F\alpha}})|0\rangle
\equiv \lim_{k\to \pm 2q_{F\alpha} } \langle \mp q_{F\alpha}^{\pm },
k;\alpha | J_0^{\vartheta }(k)|0\rangle $ \cite{Carmelo92b}.
Furthermore, $\langle \alpha | J_0^{\vartheta } (k_{\to 0})|0\rangle
=\langle \alpha | J_0^{\vartheta } (k_{\to  2q_{F\alpha}})|0\rangle
$. Therefore we have $\nu $ different elements $\langle \alpha
| J_0^{\vartheta }(\omega_{\to 0})|0\rangle \equiv \langle \alpha
| J_0^{\vartheta }(k_{\to 0})|0\rangle = \langle \alpha
| J_0^{\vartheta }(k_{\to 2q_{F\alpha}})|0\rangle $. Since
\cite{Carmelo92b}

\begin{equation}
\langle \alpha | J_0^{\vartheta }(\omega_{\to 0})|0\rangle =
\sum_{\alpha '} g^{\vartheta }_{\alpha '}\langle \alpha
| J_0^{\gamma (\alpha ')}(\omega_{\to 0})|0\rangle \, ,
\end{equation}
we conclude that the $\nu \times \nu$ matrix elements
$\langle \alpha | J_0^{\gamma (\alpha ')}(\omega_{\to 0})|0\rangle $
of the $\nu$ Cartan currents fully define the general currents
$J_0^{\vartheta }(k)$ in ${\cal H}_0$. An important result is that
the charge dressed matrix $Z^1$ represents the $\nu$ Cartan operators
$J_0^{\gamma (\alpha )}(k)$ in ${\cal H}_0$, i.e. its elements
are the above $\nu\times\nu$ matrix elements,

\begin{equation}
\xi_{\alpha \alpha '}^1 =
\langle \alpha | J_0^{\gamma (\alpha ')}(\omega_{\to 0})|0\rangle
\, .
\end{equation}
This result also holds for single-component systems with $\xi^1=
\xi_{\alpha \alpha }^1$ being the ``dressed charge'' of Korepin
\cite{Korepin}. To prove $(5)$ we use a method introduced recently
to derive two-particle matrix elements \cite{Carmelo92}.
We evaluate $\chi^{\vartheta}=\lim_{k\to 0}\lim_{\omega \to 0}
\chi^{\vartheta}(k,\omega )$, where $\chi^{\vartheta}(k,\omega )
\equiv -\sum_{l}|\langle l| J_0^{\vartheta }(k)|0\rangle |^2
[2\omega_{l0}/(\omega_{l0}^2-(\omega+i\eta)^2)]$ is the
$\vartheta - \vartheta$ correlation function (the $l$ summation is
over all the eigenstates and $\omega_{l0}$ are the corresponding
excitation energies). $\chi^{\vartheta}$ can be obtained by a linear
response analysis \cite{Carmelo92b,Carmelo92}. We find
$\chi^{\vartheta} =- {N_a \over {\pi }}\sum_{\alpha '} [\sum_{\alpha
''} g^{\vartheta }_{\alpha ''} \xi^1_{\alpha ' \alpha ''}]^2 /
v_{\alpha '}$ and $\chi^{\gamma (\alpha )} = - {N_a \over {\pi }}
\sum_{\alpha '}[\xi^1_{\alpha ' \alpha }]^2 / v_{\alpha '}$.
Since for $\lim_{k\to 0}\lim_{\omega \to 0}$ the operator
$J_0^{\vartheta }(k)$ only couples the ground state to the
pseudoparticle single-pair eigenstates of ${\cal H}_0$, these
correlation functions can alternatively be written as
$\chi^{\vartheta } = - {N_a \over {\pi }}\sum_{\alpha '}
|\langle \alpha '| J_0^{\vartheta }(\omega_{\to 0})|0\rangle |^2 /
v_{\alpha '}$ and $\chi^{\gamma (\alpha )} = - {N_a \over {\pi }}
\sum_{\alpha '} |\langle \alpha '| J_0^{\gamma (\alpha )}
(\omega_{\to 0})|0\rangle |^2 / v_{\alpha '}$. Comparision of the
alternative $\chi $ expressions and the relationship between the
currents imply the validity of Eq. $(5)$. The $\xi_{\alpha \alpha
'}^0$'s can also be expressed in terms of the matrix elements
$(5)$. The dependence of the $\xi_{\alpha \alpha '}^i$'s
on the phase shifts $\Phi_{\alpha\alpha '}^{\pm }$ reveals
that the overlap in the matrix elements $(5)$ is regulated by the
pseudoparticle forward-scattering interactions. In the case of the
$[U(1)]^{\nu }$ non-interacting phases the terms involving the phase
shifts vanish and the matrix elements reduce to the term $\xi_{\alpha
\alpha '}^1=\xi_{\alpha \alpha '}^0=\delta_{\alpha ,\alpha '}$
\cite{Carmelo92b} (in the case of higher-symmetry non-interacting
phases the $\xi_{\alpha \alpha '}^i$'s are also constant, but not
necessarily given by $\delta_{\alpha ,\alpha '}$). In turn, in
interacting 1-D Landau liquids these terms are finite
\cite{Carmelo92b,Carmelo92} and introduce interaction-dependent
matrix elements (some with $\alpha \neq \alpha '$). These matrix
elements appear in the excitation energies of the HWS through
the conformal dimensions $h^{\pm}_{\alpha}$ and therefore the
interaction dependence of the critical exponents associated with the
oscillating terms of the correlation functions is determined by the
pseudoparticle interactions \cite{Frahm,Carmelo92b}. The energy
spectrum of the excitations (C) {\it is not affected} by these matrix
elements.

Since in the interacting case $J_0^{\gamma (\alpha )}(k)$ couples
$|0\rangle $ to single-pair eigenstates of branches $\alpha '\neq
\alpha$, from the point of view of the matrix elements $(5)$ the
particle quantum numbers $\gamma (\alpha )$ and the pseudoparticle
quantum numbers $\alpha $ are related but are not fully equivalent
numbers. However, from the point of view of the stress tensor-tensor
correlation functions $(3)$ and Virasoro algebra generators
$L_{n}^{(\alpha)}$ and $\bar{L}_{n}^{(\alpha)}$, which
only involve excitations (C), $\alpha$ and $\gamma (\alpha )$
can be considered equivalent numbers. We finish this Letter by
showing that such ``equivalence'' also shows up in the stiffnesses
$D^{\gamma (\alpha )}$ of the conductivity spectra associated with
the components of the Cartan currents $J_1^{\gamma (\alpha )}$. We
want to investigate the effect of the low-energy Hilbert-space
decoupling (in $\nu $ branches of gapless excitations) on the
conductivities associated with the particle quantum numbers. We show
below that this leads to the concept of dynamical decoupling. The
real part of the frequency-dependent conductivity associated with a
current $J_1^{\vartheta }$ can be written as \cite{Carmelo92b}
${\cal R}\sigma^{\vartheta }(\omega ) \equiv 2\pi (e_{\vartheta
})^2[D^{\vartheta } \delta (\omega )+{1\over {N_a }}\sum_{l\not= 0}
|\langle l| J^{\vartheta }_1| 0\rangle |^2 \delta (\omega_{l0}^2-
\omega^2)]$, where $e_{\vartheta }$ is the elementary physical
constant connected to $\vartheta $ (in the case of electronic
systems $e_{\rho }=-e$ (electronic charge) and $e_{\sigma_z }=1/2$
(spin), for example) and the summation is over all the eigenstates.
Based on a relation between $\chi^{\vartheta}(k,\omega )$ and
$\sigma^{\vartheta }(k,\omega )$ one can derive the following
expressions for the stiffness $2\pi D^{\vartheta } = \sum_{\alpha '}
q_{F\alpha '} / m_{\alpha '}^{\vartheta }$ and transport mass
$m_{\alpha }^{\vartheta }\equiv q_{F\alpha }/[g^{\vartheta }_{\alpha}
\sum_{\alpha '}\sum_{\alpha ''} g^{\vartheta }_{\alpha ''} v_{\alpha
'}\xi^1_{\alpha '\alpha }\xi^1_{\alpha '\alpha ''}]$. The latter
represents the $\vartheta$-transport mass for a $\alpha $-
pseudoparticle of pseudomomentum $q=\pm q_{F\alpha }$
\cite{Carmelo92b}. If we choose $\vartheta = \gamma (\alpha )$, we
find $2\pi D^{\gamma (\alpha ) }=q_{F\alpha }/m_{\alpha }^{\gamma
(\alpha) }$, where $m_{\alpha }^{\gamma (\alpha ) }=q_{F\alpha }
/[\sum_{\alpha '}v_{\alpha '}(\xi^1_{\alpha '\alpha })^2]$, and
$m_{\alpha '}^{\gamma (\alpha) }=\infty $ for $\alpha \not= \alpha
'$. The pseudoparticle transport masses are the quantities which
define the low-frequency elementary currents of the many-body system
\cite{Carmelo92b}. The fact that the $\nu\times\nu$ matrix $A_{\alpha
\alpha '}=1/m_{\alpha '}^{\gamma (\alpha )}$ is diagonal in the
representation of the Cartan currents has a deep physical meaning.
The coupling constant of a $\alpha '$-  pseudoparticle to a $\gamma
(\alpha )$ applied external probe is given by $C_{\alpha '}^{\gamma
(\alpha )}=e_{\gamma (\alpha )}g_{\alpha '}^{\gamma (\alpha )} =
e_{\gamma (\alpha )} \delta_{\alpha , \alpha '}$ \cite{Carmelo92b}.
Therefore, a $\alpha '$-  pseudoparticle only couples to
$\gamma (\alpha )$ probes if $\alpha=\alpha '$, and its
$\gamma (\alpha )$-transport mass is infinite for $\alpha \not=
\alpha '$. This defines the low-energy dynamical decoupling of the
system. For example, in the HLLP one finds that while
$m_c^{\downarrow }=m_s^{\rho }=\infty$, both $m_c^{\sigma_z }$
and $m_s^{\sigma_z }$ are finite because $J_{\mu}^{\sigma_z }$
is not a Cartan current: only in the limit of zero field
$m_c^{\sigma_z }\rightarrow \infty$, i.e. we have a charge-spin
dynamical separation, as expected for the $U(1)\times SU(2)$ phase.

We thank E. Fradkin for stimulating discussions, D.K. Campbell and
P. Horsch for useful comments, and the Department of Physics of the
UIUC for support and hospitality. J.M.P.C. gratefully acknowledges
the support of Funda\c{c}\~ao Luso-Americana, University of \'Evora,
and CFMC-Lisbon and A.H.C.N. thanks CNPq for a fellowship.



\begin{references}
\bibitem[9]{Carmelo}
        Our definition of $Z^1$ is the transposition of that of
        Ref. $[4]$.
\end{references}
\end{document}